# Are Friends Overrated?

## A Study for the Social News Aggregator Digg.com

Christian Doerr, Norbert Blenn, Siyu Tang, Piet Van Mieghem
TU Delft, Mekelweg 4, 2628CD Delft, The Netherlands {C.Doerr, N.Blenn, S.Tang, P.F.A.VanMieghem}@tudelft.nl

**Abstract.** The key feature of online social networks (OSN) is the ability of users to become active, make friends and interact via comments, videos or messages with those around them. This social interaction is typically perceived as critical to the proper functioning of these platforms; therefore, a significant share of OSN research in the recent past has investigated the characteristics and importance of these social links, studying the networks' friendship relations through their topological properties, the structure of the resulting communities and identifying the role and importance of individual members within these networks.

In this paper, we present results from a multi-year study of the online social network Digg.com, indicating that the importance of friends and the friend network in the propagation of information is less than originally perceived. While we do note that users form and maintain a social structure along which information is exchanged, the importance of these links and their contribution is very low: Users with even a nearly identical overlap in interests react on average only with a probability of 2% to information propagated and received from friends. Furthermore, in only about 50% of stories that became popular from the entire body of 10 million news we find evidence that the social ties among users were a critical ingredient to the successful spread. Our findings indicate the presence of previously unconsidered factors, the temporal alignment between user activities and the existence of additional logical relationships beyond the topology of the social graph, that are able to drive and steer the dynamics of such OSNs.

## 1 Introduction

The recent explosive growth of online social network (OSN) platforms such as Facebook, Twitter, LinkedIn, or Digg has sparked a significant interest into these online platforms. As several hundred million Internet users now regularly frequent these sites as a place to gather and exchange ideas, researchers have begun to investigate how this comprehensive record can be used to understand how and why users join a community, how these networks grow by friendship relations, how information is propagated among friends, and who are the most important and influential users in such social groups. A good understanding of these principles would enable many application scenarios, such as the prediction of elections, competitions and trends [1], effective viral marketing [2], targeted advertising [3] or the discovery of experts and opinion leaders [4].

These investigations and applications in social networks however make the fundamental assumption that the friendship relations between users are a critical ingredient for the proper functioning of social networks [5], i.e., they assume that information, opinions and influences are sourced by single individuals and then propagated and passed on along the social links between members of the community. The extent, density, layout and quality of the social links and the network of links as a whole will therefore determine how information can be spread effectively.

In this paper, we report on results from a multi-year empirical study of the online social network Digg.com, a so-called social news aggregator, that indicate that the criticality and importance of individual friendship relations and the friendship network as a whole is less than previously perceived. In these social news aggregators, users submit news items (referred to as "stories"), communicate with peers through direct messages and comments, and collaboratively select and rate submitted stories to get to a real-time compilation of what is currently perceived as "hot" and popular on the Internet. Yet, despite the many possible means to communicate, interact and spread information, an analysis of ten million stories and the commenting and voting patterns of two million users over a period of four years revealed that the impact of the friendship relations on the overall functioning and outcome of the social network is actually surprisingly low. In particular, we find that while users indeed form friendship relations according to common interests and physical proximity, these friendship links are only activated with 2% probability for information propagation. Furthermore, in about 50% of all stories that became "hot", there was no prior contribution by the friend network to the extent that would have led to emerging popularity of the story; instead, we find that a critical mass was reached through participation of random spectators.

The contributions of this paper are therefore two-fold: First, we challenge the current underlying assumption in online social network research that friendship relations and the network of friendships is a critical necessity to proper information propagation in these communities and present evidence for the limited conductivity of these links. Second, we show that (a) timing and the alignment of user activities is crucial to the success of submitted news items, which can either individually or in conjunction with friendships explain the inherent dynamics, and that (b) distinct interactions patterns exist, the majority happening outside the social friendship graph.

The remainder of this paper is structured as follows: Section 2 discusses related work and prior findings on the role and characteristics of friendship links and the friendship network in online social networks. Section 3 describes background information about the social network used in our experimentation and our data collection methodology. Sections 4 and 5 present our findings on the role of friend- ships and selected individuals to the successful information propagation. Sections 6 and 7 present the role of temporal alignment between user activities and demonstrate the existence of structure in user interaction patterns outside the social graph. Section 8 summarizes our findings and outlines future research. This article is an invited extended version based on our previous conference paper [6], presented with more details and augmented with additional findings.

## 2 Related Work

Ever since the publication of Katz and Lazarfeld's argument for the origin and spread of influence through communities [7], researchers have investigated the mechanisms by which ideas and opinions are passed along social relationships. Since then, the role of individuals, as well as the characteristics and importance of their relationships have been investigated in a variety of different research fields.

A common way to describe the structure and relations between individuals in a community are by "weak" and "strong" ties. Originally proposed by Granovetter [8] in a sociological context based on the intensity, frequency and amount of personal contact between individuals, this characterization of interpersonal relationships has spread and been adopted by many other subject domains, such as marketing, political science and economics. According to the theory, weak and strong ties behave differently in communication and information dissemination: while a lot of interaction is taking place between "strong ties", i.e., persons with frequent and long-lasting contacts, these ties within tightly knit clusters carry a lot of redundant information; thus new and novel information can best enter from outside these clusters across "weak ties".

These aspects of novel information transmission and redundancy in weak and strong ties are further analyzed by Burt [9] within the context of organizational networks, who finds that information transfer in a company is best achieved when individuals possess a high number of overall, but relatively low number of redundant contacts. People switching between different positions within an organization keep their previous ties, and companies with a well-connected social network are exhibiting a larger agility to react to problems. Burt refers to areas within organizations having too few or too weak "weak ties" as structural holes. Similar findings are also reported by Krackhardt [10] who investigated the importance of informal interpersonal networks in organizations in times of crises.
In a game, two hypothetical companies were created in which the units in one company contained friends working together in one division and in the other company friends have been in different units. During crises, simulated through a drop in available resources, the organization having a well-connected network of units performed significantly better than the other one.

Hansen [11] added to this so-called search transfer problem the notion that besides the existence of weak and strong ties, the absolute strength of a connection is also of noteworthy importance. In a study of information sharing between subunits of large multi-national companies, well-connected units again scored better than others, but among equally well-connected subunits the ones with more intense ties performed even better due to increased collaboration. When sharing complex knowledge, weak connections did provide exposure and information about possible solution approaches, successful adoption however was aided by an increased intensity of the social tie.

When facing problems or difficult questions, we turn to friends or acquaintances around us to get clues or a solution, as claimed by Homans [12] and Coleman et al. [13, as cited in [14]]. Consequently, social interactions and ultimately the social network provide a fertile ground for the promotion of new ideas, information and innovation. A prime example to assess such knowledge dissemination is Coleman et al.'s study "Medical Innovation" [13], investigating whether and how a group of physicians are adopting a new drug after recommendations from their social network. Later reanalyzed in [14], Burt however does not find convincing evidence that social ties were indeed the driving force behind the adoption of the new medication, as the number of ties to physicians who had adopted the drug had no influence on whether a physician was prescribing it in turn, which should occur in a contagion process.

Tsai [15] investigated the knowledge dissemination measured in terms of innovation ability within an organizational network of 60 business units and argues that the ability to obtain beneficial information depends on the location within a social network. Individuals or groups placed and connected in the center of the network get more exposure to information simply through their topological location in the system, which in his example manifested itself in higher innovation performance.

Although the potent abilities of networks to transmitting information and relaying messages have been known for quite some time since Milgram [16] conducted his famous experiment, leading to the by now well-known phrase "6 degrees of separation", interactions between the individual and the surrounding social network have received in the recent past new and diverse attention, for example from researchers in human dynamics, public health or epidemiology. Christakis [17] for example demonstrated in a longitudinal study of some 12 000 participants that a person's risk of become obese, his ability to stop smoking or maintain happiness is to a significant extent influenced by those around him. Both benefits and risks are being propagated by social ties, and the influences are contagious between friends and friends of friends. These recent outcomes have lead to new insights and impulses in public health, for example that certain diseases are better approachable at the individual and the network level or that epidemics may be more efficiently prevented under resource constraints when first protecting the high-degree entities by vaccination in the network, who act as fast spreaders of ideas and disease. The same underlying principle of the high importance of high-degree nodes can in turn however also create a significant problem, as these hubs pose a significant vulnerability in scale-free network topologies, for example when targeted by malicious attacks [18].

In recent years, two distinct trends have emerged in network analysis: First, with the availability of new datasets and fast processing options the focus has shifted from empirical observation of small groups of a few dozen to a few hundred participants towards larger studies. With the advent and wide-spread popularity of online social network platforms, this field of study has gained additional momentum as these newly available communities now provide an easily accessible, machine-readable data source for a broad-scale analysis of established research topics.

Second, the bulk of recent work has investigated the structural properties of complex networks, with a lesser focus on understanding the friendship and information propagation processes taking place inside such large scale social networks.

Along these lines, Mislove et al. [19] studied the topological properties of four OSNs at large-scale: Flickr, YouTube, Live-Journal, and Orkut. By crawling publicly accessible information on these networking sites, they evaluated different network metrics, e.g. link symmetry, node degree, assortativity, clustering coefficient of the four networks. The OSNs investigated were characterized by a high fraction of symmetric links, and composed of a large number of highly connected clusters, thereby indicating tight and close (transitive) relationships between users. The degree distributions in the OSNs follow a power law and the power law coefficients for both in-degree and out-degree are similar, showing the mixed importance of nodes in the network - there are few well connected and important hubs to which the majority of users reach to.

In [20], Leskovec et al. presented an extensive analysis about communication behaviors and characteristics of the Microsoft Messenger instant-messaging (IM) users. The authors examined the communication pattern of 30 billion conversations among 240 million people, and found that people with similar characteristics (e.g., age, language, and geographical location) tend to communicate more. The constructed communication graph was analyzed for topological properties of the graph in terms of node degree, clustering coefficient, and the average shortest path length; it was shown that the communication graph is well connected, robust against node removal, and exhibits the small-world property.

Backstrom et al. [21] studied the network growth and evolution by taking membership snapshots in the LiveJournal network. They also presented models for the growth of user groups over time. Benevenuto et al. [22] examined users activities of Orkut [23], MySpace [24], Hi5 [25], and LinkedIn [26]. A clickstream model was presented in [22] to characterize how users interact with their friends in OSNs, and how frequently users transit from one activity (e.g. search for peoples profiles, browse friends profiles, send messages to friends) to another. There are also many researchers who aim to discover content popularity and propagation in OSNs. For instance, in [27], photo propagation patterns in the Flickr OSN are studied. The results discovered in [27] reveal different photo propagation patterns, and suggest that photo popularity may increase steadily over years. An in-depth study about content popularity evolution and content duplication was performed with YouTube and Daum (a Korean OSN) in [28]. In this paper, Cha et al. studied the popularity distribution of videos uploaded to the two websites. It was shown that video popularity of the two applications is mostly determined at the early stage after a video content has been submitted to the OSNs. A similar comparison was conducted for YouTube and Digg by Szabo and Huberman [29], who presented a model of predicting the long term popularity of user generated content items.

Besides high-level observations, there is only little known yet about the exact content propagation mechanisms taking place inside an online social network and the possible roles and impact different types of users might assume during information dissemination. If unearthed, such insights would have many application domains, ranging from the discovery of experts and opinion leaders to efficient innovation adoption and marketing. For this reason, the search for the "influentials" has been a significant endeavour in the viral marketing literature where it is argued that "few important trends reach the mainstream without passing the Influentials in the early stages, ... they give the thumbs-up that propel a trend". Their recommendation and word-of-mouth dissemination lets information spread exponentially [30, p. 124].

As it is however quite difficult to track the spread of content in online social networks and through this generate the basis to search for and study the role of influential users or influencing relationships, there exists to this date no study that evaluates the role of users and friendship within online social networks at a population scale, i.e., across millions of subscribers. This paper aims to address this void and investigate for the entire online social media aggregator Digg.com the following two hypotheses that are frequently assumed in social network analysis:

$H_1$  There exist critical members inside the community who have better or earlier access to important information.

$H_2$  Inter-personal relations and the overall network of friendships are the key component to the successful spread of information.

The remainder of this paper will dissect each hypothesis and evaluate it empirically.

## 3 The Digg data collection

The news portal "Digg.com" is a social content website founded in 2004, which according to the rating provided by Alexa Internet, Inc. [31] in 2010 belonged to the top 120 most visited websites in the Internet. At the end of the presented study, a total of 2.2 million users were registered on the webpage, submitting between 15,000 to 26,000 stories per day to the system. Out of those submitted stories, approximately 180 stories per day were voted to become "popular". The collected corpus contained the activities of users and the content of more than 10 million stories in total, 200 000 of which achieved critical mass.

Within a social media aggregator such as Digg.com, registered users are able to participate by submitting, commenting and voting on content they like or dislike. Users can send in news or blog articles, images and videos by submitting a link to the web page where the information can be found, together with a title and brief description of the media item. Entries in Digg are categorized in 10 main topics (Business, Entertainment, Gaming, Lifestyle, Offbeat, Politics, Science, Sports, Technology, World News), each further divided into a total of about 50 special interest areas. Registered users and visitors to the site can browse the collection for example by category, submission time or through a recommendation engine, thus, Digg also acts as a online social bookmarking site.

New submissions to the system are enqueued in a special section of the web site called "upcoming", where entries are staying for a maximum of 24 hours. If an item generates enough attention and positive recommending votes, an activity called "digging", within this time period, the story is tagged as "popular" and "promoted" to the "front page", which is the main home page immediately visible to anyone navigating to the Digg.com website. Thus, once promoted to the front pages, a story generates a lot of attention and traffic from registered users and casual visitors. The concentrated, sudden instream of users following the link from a promoted story is often so large to frequently overload remote web servers, referred to in the community as "the digg effect".

On the Digg website, users also engage directly with each other and can create friendship connections to other users in the network. These connections can either be one-directional or two-directional, in which case the user is either a fan or a confirmed mutual friend with another person. Fans and friends are notified by the friends interface of digg if their contact has "dugg" or submitted a story. With the introduction of the most recent revision of Digg (v4.0 was released in August 2010), this notification system was even made the default option when visiting Digg, i.e., *only* stories submitted or dugg by friends were visible to the user when browsing to digg.com (a feature called "My News"), and the default had to be proactively changed to also receive other recommendations. It should be noted at this point that the semantics of a friend in Digg (obtaining information) is certainly different from a friendship in Facebook (personal acquaintance) or LinkedIn (business contact) [32], as also the main function differs between these social networks. As this paper investigates information propagation and social news aggregators such as Digg.com focus on the exchange of information, these results are only immediately applicable to this type of OSN. To what extent these findings can be extended towards other types needs more investigation.

To obtain the most complete and representative snapshot possible, we studied different aspects of the Digg OSN, such as the friendship relations, the characteristics of users, and the properties of the published content. While most social network traces are crawled using friendship relations, e.g. [19] and [33], the Digg dataset was obtained by a simultaneous exploration of the network from four different perspectives, as shown in figure 1. By using the Digg Application Programming Interface (API) and direct querying of the website, we were able to explore the aforementioned four perspectives (from bottom to top in Fig. 1) during data collection:

- **Site perspective**: The Digg website lists popular and upcoming stories in different topic areas. Every hour, we retrieve the frontpages with all popular stories (for all topics) that are listed on Digg. Every four hours, all upcoming stories (for all topics) are collected. All discovered stories are added to an "all-known story" list maintained by us.
- **Story perspective**: For each story that has been retrieved, a complete list of all activities performed by different users (who digged on the story) is collected. Any user that is discovered will be added to the "all-known user" list for future exploration.
- **User Perspective**: For each user discovered within the Digg OSN, the list of their activities, such as submitting and digging on stories, is retrieved. Occasionally, a previously unknown story is discovered (this is typically the case for older stories before we started the collection). For such a story, the entire (digging) activities of users are retrieved for that story.
- **Social Network Perspective**: Each registered user can make friends with other Digg users. In the crawling process, a list of friends is retrieved for every user. If a friend is a previously unknown user, this user is added to the data discovery process, and a list of all his/her friends and his/her public user profile information are retrieved. This procedure is continued in a breath-first search (BFS) manner until no new user can be found. The process is periodically repeated afterwards to discover new friendship relations that have been formed after the last crawling pass through the data.

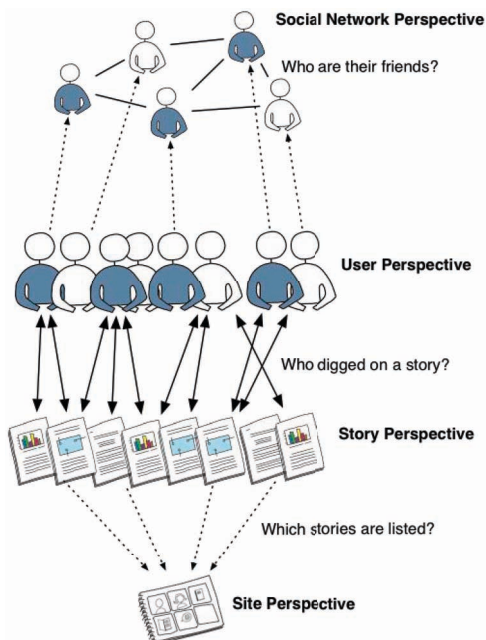

**Figure 1** *The four components of the Digg crawling process. To obtain a comprehensive picture of the activities in the Digg OSN and avoid structural omissions such as the activities of unconnected users, the network was simultaneously explored from four different angles: 1) Which stories are listed?, 2) Who digged/commented on a story?, 3) Which stories did a person submit, digg or comment upon?, and 4) Who are the friends of every known user? Newly discovered items at one level automatically fed back into the other discovery processes.*

By using the above crawling methodology, we are able to collect nearly the entire information about friendships and activities of users and the published content in the Digg network. This is a significant and important distinction as traditional crawling techniques exploring a social network based on the friendship graph will only discover those users which are engaging in active community building and are also part of the (giant) connected component of the social graph. By exploring all four dimensions simultaneously, our data collection was able to identify any user that was either (a) digging or commenting on a story, (b) submitting a story, or (c) made at least one friendship with any other user (even outside the connected component). A comparison indicated that this extended methodology was able to find nearly twice as many users than when only crawling the giant component of the social graph, and could already provide some explanation to the contrary findings presented in our paper. Until July 2010, the Digg dataset has a volume of more than 600 GB (Giga Bytes), containing the related information of about 2.2 million registered users and 12 million published stories in the Digg OSN.

# 4 Information Spread through the Network of Friends

As discussed in the introduction and related work, it is commonly assumed that the friendship relations within a social network are the critical ingredient to the successful spread of information. This section will dissect this process and investigate for the case of the Digg OSN, whether the propagation of news is indeed the result of the activation of user ties.

## 4.1 Self-Organization of the Friendship Network

According to sociological theory, friendship relations in OSN grow directed by common interests and tastes [32]. Within the Digg social network, all news stories are classified within eight major topic areas, further subdivided by 50 special interests. When matching the users' concrete digging behavior with the topic area a story was classified in, we find that the subscribers exhibit quite strong and distinct preferences and tastes for individual topic areas: Even when following the content published in several genres, most of their attention is focused on a few areas. As shown in figure 2, if a particular user reads, diggs and is therefore interested in two distinct topic areas, say for example "Science" and "Technology", almost 70% of all consumed stories fall within the most preferred genre. For three subscribed topic areas, say for example "Lifestyle", "Business" and "Entertainment", the ratios drop to 65%, 25% and 15%, thus the most preferred topic still attracts on average nearly two thirds of all clicks. Even for users interested in eight categories the top two will on average account for 60% of read stories.

Since the relative preferences between categories are quite pronounced and stable, these ranks of user interest provide a direct measure of how similar the tastes and preferences of users in their information acquisition are. When comparing two users and their ranking of topics, the number and distance of permutation steps required to transform one list into the other (the Wasserstein rank distance [34]) will act as a measure of user similarity, e.g., two identical lists will rank as 0, the same lists with the first and fourth topic exchanged as 3. A network-wide analysis of the similarities between friends shows that users directly connected to each other have a very high alignment of their preferences and tastes: 36% of rank lists are identical, 20% require one transformation, and within three transformation steps 80% of all friendship relations are aligned. People acquire and maintain friendships based on whether these future friends have previously demonstrated a similar taste and composition in their digging behavior.

Interestingly, the rate at which a user initially acquires friends and diggs on stories seems to be related to the overall lifetime of the user's account, determined as the timespan since the first registration until the last action performed by this account. Visitors who sign up and immediately form a lot of friendship relations within their first day but slow down on their second, typically abandon their profile after one week or less. The slower and more continuous friends are added to a profile, the longer a person continuous to participate on the Digg website, as shown in figure 3. The most sustainable rate of digging activity and friendship acquisition is exhibited by those who remain active for 3 years and thus can be considered heavy users of the platform.

## 4.2 Incentives for Common Diggs

While there exists a perfect overlap between the interests and tastes of individual friends, there is a surprisingly low amount of common activity among friends and on average only 2% of all friend pairs actually do react and digg on the same story.

The hypothesis that common interests result in the formation of friendships in order to gain information from neighboring peers [33] would also predict that the more similar the tastes between friends are, the closer the alignment of clicking patterns would be. In practice, we found this however not to be entirely the case; although there is a generally decreasing trend between interest overlap and common clicks, the differences are not statistically significant.

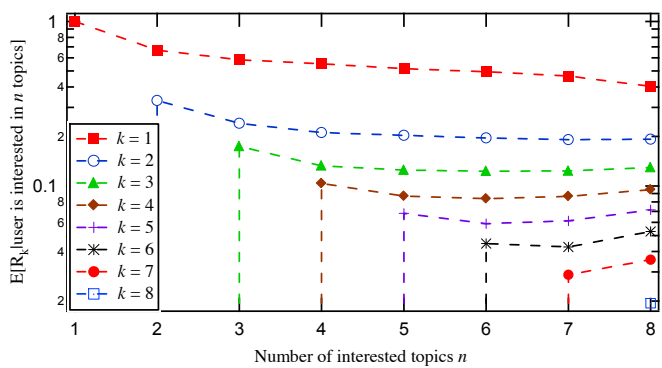

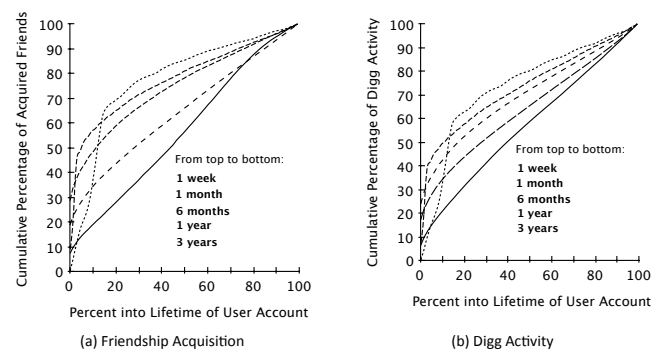

(a) Friendship Acquisition    (b) Digg Activity

**Figure 2** *The figure plots the share of diggs a user devotes on average on the $1^{st}$, $2^{nd}$, … $k^{th}$ most frequented topic areas (y-axis, in logarithmic scale) as a function of the total number of categories a user has been active in (x-axis). On average users focus the bulk of their attention on a limited number of most preferred categories. As users' interests become more widespread, this diversification comes mostly at the expense of already lesser read areas while the top choices remain relatively stable: even when reading stories from eight categories, an average user still focusses nearly 60% of all activity on the two most preferred subject areas.*

**Figure 3** *Acquisition of (a) friends and (b) diggs through the lifetime of user accounts. The y-axis shows the cumulative percentage of digging/friending activity to date as a function of the cumulative lifetime of a user account on the x-axis, defined as the total timespan between registration and the last activity of a particular account. On average the faster a user acquires friends and clicks in the beginning, the more likely the account will be abandoned after a shorter time.*

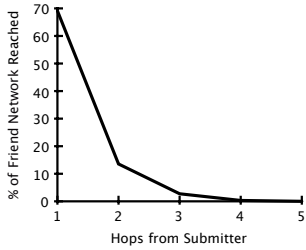

**Figure 4** *Activation of the friendship network in information spread. The figure shows the percentage of the total activated friendship network (y-axis) as the information spreads out hop-wise from the original submitter (x-axis): most of the entire friendship network is actually already covered in the first hop, beyond three links separation the activation drops to a fraction of a percent.*

### 4.3 Activating the Friends of Friends

Friends and friendship pairs however do not exist in isolation, but are embedded within a larger network of the friends of the friends. This in OSN very dense structure [19] may work as a powerful promoter, as theoretically a large number of nodes can be reached if information can be passed on from friend to friend and propagated over several steps: In theory, an information may reach an exponentially growing number of recipients as the number of hops it traverses increases. Given that there exists a critical threshold that needs to be met to promote a story to high popularity and a limited number of friends are actually active on the site on a particular day, the *network* of friends, in other words the friends of friends, could make the difference between stories that spread or fall into oblivion.

Our analysis shows that information can indeed travel over multiple hops from the original submitter in the Digg OSN (see figure 5(a)) and on average does reach 3.7 hops from the source until the propagation dies down. The actual contribution of the multi-hop network, i.e., the amount of friends of friends that can actually be activated by this process, is however rather limited. As shown in figure 4 nearly 70% of the ultimately participating *network* of friends consists of the submitter's direct contacts, while the incremental benefit of the additional hops decreases exponentially. This result is not astonishing given the generally low activation ratios of friends and possible redundancies in the spread as indicated by the dashed line in figure 5 (a), i.e., a person receiving several notifications from various friends in the previously activated friendship network.

This aspect is further visualized in figure 5(b), which shows the share of the total redundant notifications observed at a particular distance from the original source. A notification can be classified as redundant if a particular user has been informed about a particular story before and the incoming trigger consequently provides no additional information, or if a notification arrives after the receiving user has already digged on a story earlier on. As can be expected, due to the tree topology of the first hop friendship network, no duplicate notifications are initially generated, while the number of redundancies increases rapidly as the spread progresses, both as a result of back-links into the already explored network and due to exhaustion of the pool of possible candidates. The slope of both curves shown in figure 5(b), the redundant notifications within the activated friendship network indicated by the dashed line and the theoretical maximum of redundant notifications if all friends would react to an incoming trigger indicated by the solid line, is however bounded: the former one declines with a dwindling network activation after three hops, the latter one slows down as the network and all friendship links are getting saturated.

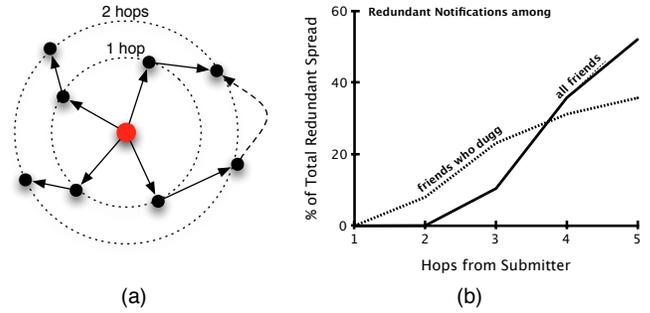

(a)          (b)

**Figure 5** *Redundant activation in the friendship network. As the friendship graph contains common acquaintances, a spreading process along the friendship links will result in duplicate, redundant notifications as indicated by the dashed arrow in subfigure (a). Subfigure (b) quantifies the percentage of the total redundant notifications passed along friendship links (y-axis), depending on the stage in the spreading process (x-axis). The plot distinguishes between the actual renotifications observed for stories (send out by those who dugg on it) and the theoretical maximum of redundancy that could be reached within the friendship network (if all triggered friends would actually digg). Although in principle growing exponentially, both curves are bounded, the former one by the dwindling activation ratios after three hops, the latter one by the near complete saturation of the digg network after five hops.*

### 4.4 Reaching Critical Momentum

All news stories submitted to the Digg social network are initially collected in the "upcoming" list, which with more than 20 000 submissions per day has a very high turnover rate (more than 800/h) and a total capacity of 24 hours after which stories will disappear. In order to become promoted to the frontpages, a story therefore has to attract sufficient interest, i.e., a large enough number of diggs, within this timeframe of 24 hours. As shown in figure 6, the majority of stories that passes this threshold does so after the initial 16 hours. We experimentally determined that about 7 diggs per hour are necessary to qualify for the promotion, thereby stories should gather on average around 110 diggs.

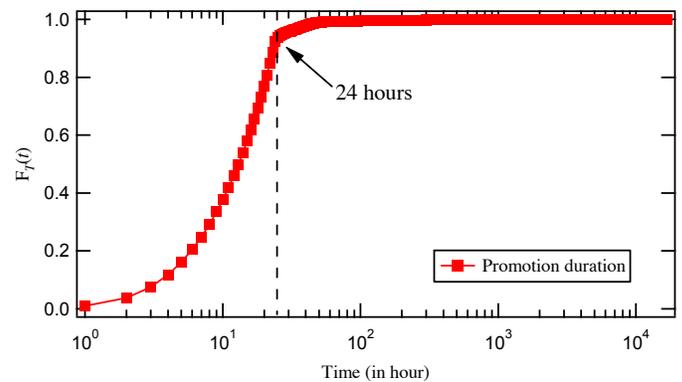

**Figure 6** *Promotion probability of submitted stories over time. Stories have to gain initially enough momentum within 24 hours to be selected from the pool of fast-moving submitted news items. The figure shows the cumulative probability for a story to become promoted within a particular time period after original submission.*

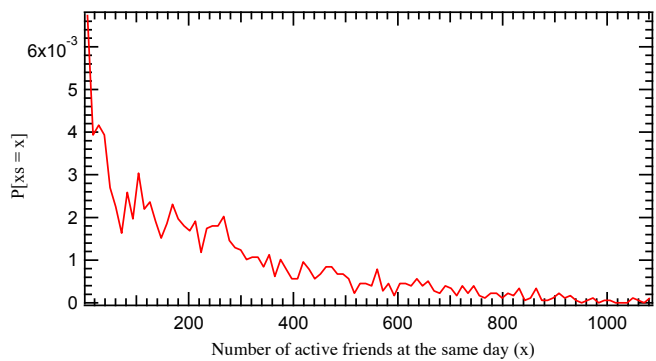

**Figure 7** *The figure displays the probability (y-axis) for a given number of friends (and friends of friends) to be online within same day (x-axis). The likelihood that a high number of friends are active on the site on the same day decreases rapidly, thus whether a story can be successfully spread just by the friendship network alone heavily depends on the performance of the underlying stochastic process.*

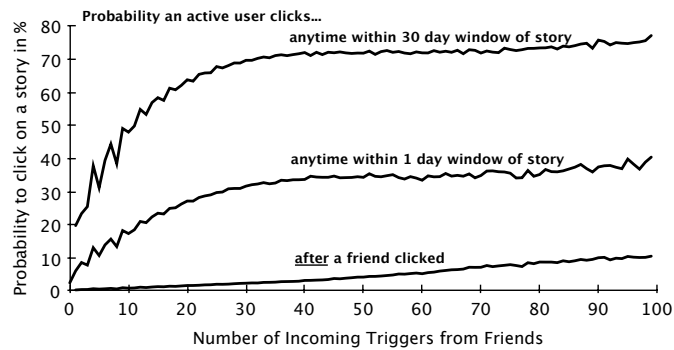

**Figure 8** *Probability of user activation after triggers. The likelihood for a user to digg on a story (y-axis) in principle increases with the number of diggs performed by the friendship network (x-axis). The effect however is drastically limited when only considering a window of 1 day, the time from submission until the promotion cut-off date when a user's contribution will have the most effect, instead of the total 30-day lifetime of a promoted story. Additionally, when also enforcing the requirement that notifications are strictly arriving before the receiving user has digged, the probability to digg even after a high number of friendship network votes drops to less than 10%.*

A story can rally this support initially from random spectators or friends of the submitter, who were notified about the newly placed story. To successfully spread via friendship links, a critical mass of friends needs to vote on the item. For this to happen however (assuming that all or a high percentage of them will react to the incoming notification), a sufficient number of friends first need to be active and active on the Digg.com website within this promotion window to become aware of the story and be able to contribute to its promotion. This probability can be inferred from previous records, as the data set contains all instances when a particular user submitted, digged and commented on stories or created friendship links since account registration. Thus, an analysis of the combined actions of a particular user provides a lower bound\footnote{as the user could have been active without having been logged in or visited and seen the website without performing any action visible in the logs} of that person's probability to be active during a typical 24 hour time window. Combining such estimates with the structure of the submitter's friendship network provides an approximation of the probability that a particular number of friends are active during the promotion window.

Figure 7 shows the average likelihood for a given number of friends to be active on the website on the same day, and therefore in theory be available to provide the required support. While the probability that the required 110 friends are indeed present corresponds with the actual promotion success ratio of 0.01, this fine line between failure and success strongly depends on the performance of the underlying stochastic process, whether at a certain time a sufficient number of friends are online and willing to support the story. In the remaining 99% of the cases, additional support needs to be rallied from users outside the submitter's friend network.

### 4.5 Are Users following the Herd?

As the impulse of a user to follow a friend's previous action is relatively low, it might simply be that more than one trigger event is needed to activate a user. There exists an established body of literature on behavioral mimicry [36], indicating that people are subconsciously copying the behavior of those around them; for example, it has been reported that the likelihood for a person to buy a computer is influenced by how many computers are owned within that person's neighborhood [37].

As our data set contains both the social relationships and actions of users, we can use this combined information to quantify to what extent such network externalities are indeed influencing the behavior of individual users, i.e., does a person's likelihood to recommend some content depend on the number of friends that have previously reacted positively to a particular item? For this behavioral mimicry to unfold, a chain of conditions however needs to be met: (1) There needs to be some mechanism that lets a person learn and observe the behavior of those around them. (2) The person needs to be active and able to receive and perceive the surrounding triggers. (3) A trigger needs to be timed in such a way that it can serve as an influencer to a person's behavior.[1] (4) If possible, a causal relationship between trigger and action should be established.

In the case of Digg, the activities of users are publicly visible to everyone, and by establishing a friendship link users can keep track of their friends' activities through notifications. As friends might not be able to receive and view such notifications due to abandoned accounts, extended absences or non-aligned activity periods (an issue further discussed in section 6), only those users will be considered during the analysis that were active at least once on the Digg website during a particular story's life time, and thus could in theory have received triggers resulting from their friends' activities.

For these generally active users, figure 8 shows the probability that a person will click on the same story as one of their friends, depending on the total number of triggers received through their following relationships. As can be seen in the figure, this likelihood significantly increases with the number of incoming notifications, but saturates beyond 40 triggers. The overall activation level however highly depends upon the time frame of observation: If a user may react anytime within a 30 day time window, given enough triggers users on average click nearly on 75% of those stories as their friends have done before. This 30 day time window is however the maximum lifetime of promoted stories, which with 1% of all submitted stories (~ 160 daily) only encompass a small fraction of the overall news content on the site. New submissions have to reach the promotion threshold within 24 hours and for this time window the maximum saturated activation probability across all stories drops to about 35%.

---

[1] For the case of the adoption of computers as presented in [37] for example, the person should have bought a computer after those around it have done so. If for example the computer has been ordered months ago but not delivered yet, intermediate purchases from friends and neighbors could not have served as a trigger to that person's decision.

The presented activation ratios (and many of those studied in the previous literature) have so far only looked at a user's individual behavior and the existence of a social relationship, in other words we have not yet utilized any temporal information that can help answer whether the flow of information was aligned in such way that the incoming trigger could indeed have initiated the resulting behavior by the follower. If we make this distinction and only consider any diggs that have been made on a story *after* a friend has digged on it, the probability of a reaction drops below 3% for a low to moderate number and below 10% for even 100 incoming notifications. Given that an active user receives on average 13.7 triggers, this further explains the low conductivity of friendship links, and the resulting linear relationship between number of triggers and digging probability can be reduced to a stochastic counting process [38]. It can be expected that the percentage of *causal* triggers will even be considerably lower; to establish this number however direct feedback from participants would be necessary explaining the motivation for every digg.

### 4.6 Promotion without Friendships

The fact that the likelihood that a story can become popular solely through the activity of the submitter's friendship network is rather slim (given the slow activation ratio of friends, the limited contribution of the network of friends and low probability of a sufficiently large critical mass of friends that are active on the same day), in most cases the contribution of non-friends is necessary to promote a story up to the threshold level.

When analyzing the ratio of clicks from friends in the submitter's network to the total number of diggs before reaching the promotion threshold, the body of stories can be divided into two distinct groups - one with a high average contribution of friends and one with a low average contribution. Table 1 shows the ratio of friends and non-friends active on a story both before and after the promotion for all stories that became popular within the Digg network, divided into two groups using the arithmetic mean of friendship contribution ratios of popular stories with a friendship contribution (50%) as a dividing threshold. Stories with more than 50% friendship network contribution were tagged as (a) "friend promoted", with less than 50% as (b) "non-friend promoted". Although being a rather simplistic decision point, it provides a rather pronounced differentiation of all stories into two groups.

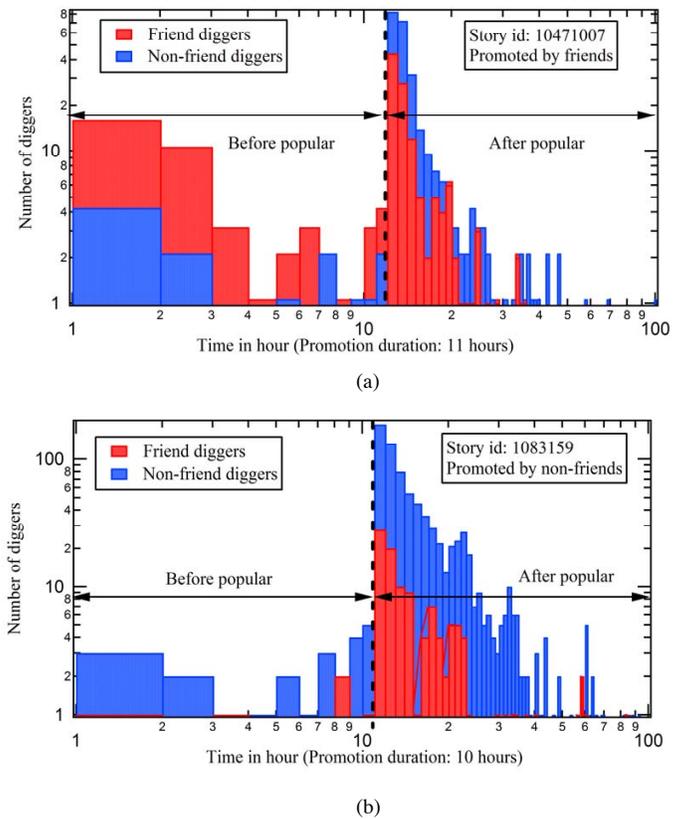

**Figure 9** Comparison of friend/non-friend digging activities over a story lifetime. The figure shows the activity patterns on a popular story since publication, where subplot (a) depicts a typical friend-promoted story and subplot (b) a typical non-friend promoted story. The x-axis lists the time in hours since story submission in logarithmic scale, the y-axis the number of diggs during this particular hour in logarithmic scale broken up in the contribution of friends (red) and non-friends (blue). Friend-promoted stories get a significant initial push from users in the submitter's friendship network, while this contribution is absent for non-friend promotions. After reaching the promotion threshold (dashed line), both stories attract a large amount of attention and recruit part of the remaining friends who were not previously activated.

In about 54% of all cases, a story was marketed predominantly by friends, although a contribution of non-friends (28%) was necessary until the story reached critical mass. Figure 9(a) shows this aggregated pattern for a prototypical story from this class; in the beginning of the stories' lifetime, the submitter's friends dominate the process until about one hour before the promotion is reached, a number of unrelated users push the story over the threshold. In the remaining cases (46%), stories were spread and consumed predominantly by users outside the submitter's friendship network. Figure 9(b) shows a prototypical example for this pattern. Once the promotion threshold is crossed, both types of stories are read more by non-friends, as the quantity is usually significantly larger and the possible contribution of the submitter's friendship network may have already been exhausted. At this time stories also experience an immediate and drastic boost in the number of incoming diggs due to the prominent placement at the top of the Digg home page. This effect however quickly dampens down again as other more recently promoted stories displace the item from its prime position and the story moves on to later front pages. Experimental measurements have determined that stories attract practically no notable number of diggs after front page 4 or 36-48 hours.

**Table 1** Ratio of friends and non-friends among the total number of diggers for popular stories. The table lists the share of diggs coming from the submitter's friendship network out of all diggs, both before and after reaching the promotion threshold. When stories are classified into (a) friend promoted (>50% friend influence) and (b) non-friend promoted, for (a) the distinct influence of the friendship network in the promotion becomes visible, where stories in (b) were dependent on a contribution from non-friends to go viral.

|  | Before popular | | After popular | |
|---|---|---|---|---|
| Average ratio | Friends | Non-friends | Friends | Non-friends |
| a) friend-promoted | 0.72 | 0.28 | 0.25 | 0.75 |
| b) non-friend promoted | 0.23 | 0.77 | 0.14 | 0.86 |

# 5 The Criticality of Individuals

As shown in the last section, the successful spread of information cannot be explained directly from the social ties inside the investigated online social network, neither through the relationships among individual friends nor from the usage and outreach of users into their friendship network, in other words activating the larger body of friends of friends. In both cases, the average activation of users is generally too small to cross the threshold to criticality, thus resulting in the fact that only 1% of all stories and items submitted to the network ever reach popularity and in only 50% of the popular stories this promotion is due to the action of friends. This however naturally raises the question whether all users are equal inside the network, or whether there are some individuals in the social community (a) who themselves have better (or earlier) access to important content and are therefore able to submit a high number of stories that will become popular, (b) can use their friendship network more efficiently, act as motivators and are able to over-propotionally recruit friends to click and spread the word, or (c) are able to early on spot content that will later resonate with the masses and become a hit. These questions will be the focus of this section.

There exist a number of ways to define the importance or criticality of individuals in networks. In complex network theory and social network analysis, importance is typically defined from a structural perspective, using topological metrics such as node degree or betweenness [39], which measure how well a particular node is connected to its surrounding peers and how many theoretical communication paths between nodes in the network will pass this entity en route.

Based on this definition, most studies of online social networks find a small number of topologically critical nodes [19, 22, 20], resulting from the typical power-law degree distribution of these complex networks; there exist a few well-connected nodes with whom a large number of users are friends. In our analysis, we confirm these findings and will thus for now use this definition and this selected group as the reference to study critical individuals.

**Table 2** *Fraction of symmetric links in the Digg network. The likelihood to reciprocate incoming social ties and turn followers into bi-directional friends decreases with the total number of followers a particular user has. For accounts with less than 10 connections more than half of all ties are bi-directional, while the highest-connected nodes in the network only maintain a mutual friendship with less than a third of their connections.*

| Degree of Node | Number of Users | Symmetric Link Ratio |
|---|---|---|
| $0 < D_{in} < 10$ | 282536 | 0.53 |
| $10 <= D_{in} < 100$ | 49416 | 0.42 |
| $100 <= D_{in} < 1000$ | 13993 | 0.39 |
| $D_{in} = 1000$ | 111 | 0.31 |

Contrary to other online social networks however, we do not only observe a skewed distribution in the degree and connectivity of nodes, but also in the symmetry of relationships among users. While most OSN show high levels of link symmetry[2], for example 74% of links in LiveJournal and 79% of links in YouTube are found to be bi-directional [19], the relationships in Digg are less reciprocative (38% on average) and also vary with the degree of the node: the more connections an individual *B* already has, the less likely it is to match an incoming new friendship request from *A*. In Digg, *A* thus becomes a "fan" of *B*, thereby receiving notifications about the activities of *B*, but this link and propagation of information remains unidirectional as *B* will not be informed about the actions of *A*.

This finding is consistent with sociological theory and ethnographic studies of social networks which identified that friendship requests in OSN are often driven by users' interest to become passively informed by means of these social ties [35, 40]. The fact that the average symmetry is significantly lower and also dependent on the degrees of remote nodes, underlines (a) that users are engaging in friendships in the Digg OSN with the intention of information delivery and (b) the existence of individuals which act (or views themselves) as sources and broadcasters of knowledge, which according to [30] would embody the critical influentials in the network.

## 5.1 Submitting Successful Stories

When looking at the entire body of stories submitted to the social news aggregator in the past 4 years, similar patterns of varying importance become visible. While a large number of people is watching the content published on Digg[3], only a limited number of registered users are actively submitting content to the social network. The activity patterns of these users is furthermore biased, as shown in the Lorentz plot in figure [10]: the 80% least active users of the network are together submitting only about 20% of the entire content as shown by the dashed red line. This indicated a very uneven and biased system[4], nearly the same skew - commonly referred to as the 80-20 rule - has been found repeatedly in economics and sociology.

This skew becomes more drastic when only considering those stories that gained enough support and were promoted to popular items. As the figure shows, these successful stories can be attributed to a select minority of only 2% of the community, which is able to find and submit 98% of all stories that will go viral. This effect is however not the result of the pure quantity that users participate in the story submission process, in other words there exists no statistically significant relationship between the number of stories a person has submitted and the ratio of stories that will become popular ($r^2$ =-0.01).

While the presence of such a highly skewed distribution pointing out a few users might indicate the existence of a few "chosen ones", a closer inspection reveals that these highly successful submitters are not those users critical for the effective spread of information. First of all, the average ratio of popular to submitted stories of the top 2% successful members of the community is only 0.23, therefore, even though they are the submitter of eventually highly popular content, they do not always generate top hits but a high proportion of their submitted content will not reach far. Second, the group of users who rank among the top successful members of the community is highly volatile. When comparing the top submitters between adjacent months or quarters, the set of successful users changes substantially between each studied time interval. As we do not find a significant number of stable members who are able to continuously repeat their previous successes, it has therefore to be concluded that there exists no conceptual difference or strategic advantage with those who do score successful stories. It appears that they were simply in the right place at the right time.

---

[2] If user *A* names *B* a friend, *B* also refers to *A* as friend.

[3] A combined analysis of user comments, diggs, and the number of visitors that followed a link associated with a particular story indicated that per registered digging user, the content is additionally seen by 12.9 passive spectators. The topics and generated clicks between spectators and digging users also reveal a near perfect overlap between digging and identified reading and usage patterns ($r^2$ = 0.96), thus the registered digging users may be viewed as a true proxy for the behavior of the entire Digg population. When combining these two user groups and their clicking behavior, we were able to account for more than 95% of the page hits referrals Digg.com is generating in the Internet according to [31].

[4] An unbiased, equally balanced population is described by the "line of equality", where the top k% of users would contribute exactly k% of the content.

We can however confidently say that it is not predominantly the well-connected nodes that are the originator of wide-spreading content, as there is no significant relationship between a user's success ratio and its level of connectivity with those around it ($p>0.5$).

## 5.2 Activation of the Social Network

While there do not exist any particular nodes that are over-proportionally injecting popular items into the network, there is the possibility that these nodes are highly successful in activating their surrounding friendship network, and therefore would be a key component in helping either their own or a friend's story reach widespread popularity.

It turns out however that the activation ratio of a node's direct friends is surprisingly low. On average, a particular node is only able to generate 0.0069 diggs per friendship link. This is mainly due to a combination of the already low conductivity of friendship links with low activity cycles of users. This low level of recruitment is furthermore quite stable with the structural properties of the network nodes. While the literature predicts that nodes in a social network achieve an exponentially increasing influence compared to their own importance [30, p. 124], we find a solid linear relationship ($r^2 =0.76$) between the size of a nodes' friendship network and the amount of users a person can recruit to click on a story, and a low slope of the linear regression ($a=0.102$). In consequence, there is no over-proportional impact of higher-degree nodes: 1 activated user with 100 friends is on average about as effective as 10 activated users with 10 friends.

While we find no *quantitative* difference in the friendship network surrounding the important nodes, there may be a *qualitative* difference in terms of structural characteristics and the information propagation along links. As complex networks evolve, certain growth processes such as preferential attachment [41] create sets of highly connected clusters, which are interconnected by fewer links. According to social network theory [8,42], these links among clusters, commonly referred to as "weak ties", act as a critical backbone for information propagation, as information within a cluster is communicated and replicated between nodes thereby creating high amounts of redundancy, while the weak ties transport other, previously unknown information between groups of nodes (see solid vs. dashed lines in figure 11(a)).

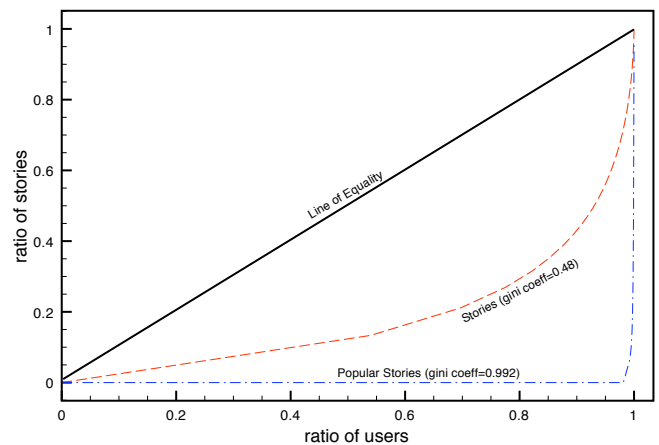

**Figure 10** Equality of story submission. The figure shows a Lorentz curve of the total and popular story submission compared to the Digg user population, i.e., the plot shows on the y-axis which ratio of stories y% were submitted by the least active x% of the Digg users. A perfect equality of activities is characterized by the line of equality, where k% of users are contributing k% of all stories. Within the Digg website, the story submission process is however unbalanced, as the least active 80% of users only submit about 20% of all stories on the Digg website. This bias is even stronger when only considering those stories that reach popularity: the 2% most successful users contributed nearly 98% of all stories that reached the promotion threshold.

To evaluate this hypothesis, we classified the network structurally into weak and strong ties according to their edge betweenness and compared their theoretical importance to the actual amount of content that was propagated between each two nodes. Figure 11(b) shows a Lorentz plot of the link weight distributions for the topological betweenness and the actual information conductivity, demonstrating that the distributions are in general comparable and of the same class. As there is no hard threshold for what characterizes a weak or strong tie, we classified the top and bottom 20% of the distribution as weak and strong ties respectively and compared them to the number of stories propagated along a certain link. As shown in figure 11(c), there does not exist any relationship ($r^2 = 0.00006$), thus information is not propagated more effectively along weak ties. Other topological definitions of how central a user is within a network, such as coreness or eigenvector centrality, also do not show any significant relationship to the propagation of information along edges ($r^2 = -0.0112$ and $r^2 = -0.0116$, respectively).

## 5.5 Early Predictors

Finally, we investigated if the assumed critical individuals -- while not able to submit more popular content or activate more users -- are able to early-on identify content that will later on become popular (see for example [30]). In the months of April-May 2009, we followed the voting patterns of all registered users on all stories to determine how successful users were in finding and clicking on content that within the next hours or days would reach the popular stage. Of all activity within this two month time period, users identified and reacted on average only to 11.9% (9% when eliminating those users who clicked on less than 5 stories in total over the period of two months) of content before it got promoted. With the absence of any high performers, we are unable to identify specific individuals who are able to consistently and repeatedly find emergent trends.

This observation did not change either for the case of the high degree individuals or the users with a high success ratio of submitting content that will go viral; there exists no statistically significant difference in their ability to find content in the social network before it actually reaches widespread popularity.

# 6 Beyond Static Friendship Relations

From the previous discussions in sections 4 and 5, it becomes evident that neither the importance of individual users nor the dynamics of the individual friendship relations or the network of friends can at a statistically significant level consistently explain why a certain story will become a success while another one will not. Furthermore, as in nearly 50% of all stories the promotion process took place without any dominant interference by the friendship net- work, we further investigated how the low participation values of the friendship network may be explained and which features are the dividing force between those stories pushed by friends and those promoted by the general public.

## 6.1 Spread Without Friends
## - A Matter of Timely Relevance

To get to the root of why one story is propagated by the help of friends while another one is pushed by random users from the community, we conducted a survey and presented a group of non-experts with the title, description, image and the type of story (news article, video, or image) of the 158 most successful stories that were promoted in the last year. As we could in retrospect classify these stories as friend or non-friend promoted, the survey items were balanced in terms of topic areas to mimic a similar distribution as on the Digg website. Given only the contextual information about the story, we asked the participants to rate the general appeal, their own personal interest and the general importance of a particular story. Using a similar representation as on the Digg website, one story was presented at a time to the participants to rule out any influences from adjacent items or possible other cognitive biases such as the primacy effect [43]. The survey results indicated that the differentiation in the promotion process of stories was a direct result how important and relevant the participants rated the topic of a particular story. Either a high rating of "general interest to the public", in other words it is likely that one would hear about the topic in the evening news, or a high level of timely relevance, i.e., will this story be as important next month as it is now, was able to serve as a reliable predictor that a particular story has reached popularity on its own without driving help of friends (both factors statistically significant at p=0.05).

## 6.2 Explaining Critical Mass Through Temporal Alignment

As a large number of factors previously hypothesized to be of critical importance to information spread in OSNs turned out to be rather insignificant and furthermore highly volatile between observation periods, we further investigated the influence of time on the story propagation process. We found that some of the unexpected low or highly fluctuating factors are to some extent dependent upon the temporal alignment of users, i.e., whether users in general (and friends in particular) are visiting the site within the same narrow time window or not.

Figure 12 visualizes this idea of temporal alignment on a snapshot of the frontpages from April 2009, which shows the position of all popular stories with at least 100 diggs over a 48 hour time interval on the first 50 frontpages. As can be seen from the figure, there exists a high flux in the amount of stories passing through; within on average 3 hours the entire content on the first frontpage has been replaced by newer items. From a combined analysis of voting patterns and such frontpage traces, we are able to determine the usual search strategy and depth of users inside the social network, i.e., when, how often and how deep they are looking through the entire site. This process revealed that stories accumulate 80% of the entire attention they will receive after promotion from users on the first and second page only, while the ratio of users who are going over more than the first 4 front pages is practically zero.

Considering the case of two users active on 20.4.2009, this can explain the surprisingly low amount of common friendship activations, as nearly 70% of the stories visible to user A during the two morning visits are already outside of user B's attention window as the user visits the social network just six hours later. Unless B actively looks for and follows up on A activity, the abundance of content and high turnover rate of information combined with limited attention span will therefore largely bury the potential for commonality unless users proactively follow up through friendship relations. This finding demonstrates that whether a story reaches critical mass depends to a significant extent upon who and how many people are currently active on the site within a short time window. A combination of this temporal perspective with interest and friendship data can go a long way, as we were able to improve our analysis accuracy of the activation ratio of certain friendship links and parts of the friendship network by a factor of 15. Note however that while a temporal view is currently able to reveal in retrospect why certain users clicked on a particular story and how and along which parts the information did propagate, it is not yet possible to predict how users will interact on a story in the future for a variety of reasons. Most importantly, an accurate prediction will require a good model of users' future activity periods at a fine enough resolution to minimize the prediction error of which stories users will see. Further- more, it will necessary to further understand the concrete decision process that will lead to a user actively clicking on a story.

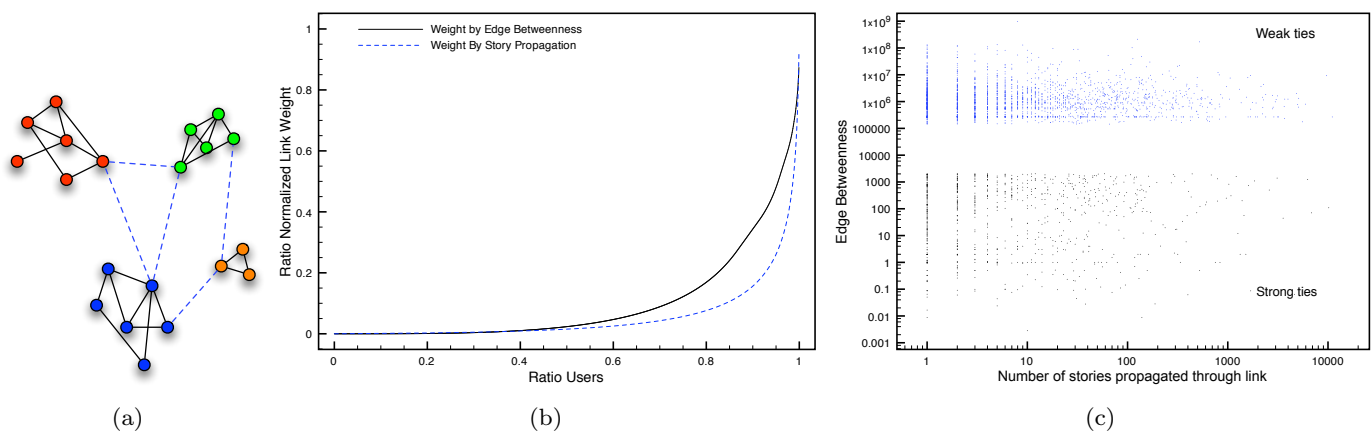

**Figure 11** Information propagation along weak and strong ties. According to the weak ties hypothesis [8,42], the links connecting different clusters and communities (resulting therefore in a high edge betweenness) are critical to the spread of information (see subfigure a). A comparison of the topological structure and the activity and usage patterns of the social network through Lorentz curves showed in principle similar network characteristics (subfigure b), yet there existed no relationship between the strength of the tie (edge betweenness) and the amount of information propagated, neither for the entire network as a whole nor for the subclasses of strong and weak links (subfigure c).

# 7 Beyond Static Friendship Relations

While critical mass can be significantly better explained when accounting for time differences and the shift and alignment of user activity periods, the individual friendship relations and the network of friends of friends still cannot fully describe the information propagation processes observed in the Digg social network. This section will present the case that a social network can only be partially captured through the topology of the direct *friendship network*, but that there may exist an unknown number of different *logical network layers* on top, whose topologies may reveal where and how interaction and collaboration actually take place.

## 7.1 Assessing the Impact of the Topological Layer

In order to discover patterns of people and groups of people who commonly act together instead of only those who seem connected through a friendship relation, we analyzed the corpus of diggs for the existence of association rules, a machine learning technique which has previously provided merit in software debugging and marketing [44]. Association rules capture and quantify the co-occurrences of particular entities, i.e., they discover if for example whenever a feature A appears, in how many cases feature B would co-occur. It has been frequently quoted that this technique has provided input to shopping center optimizations, discovering unknown, hidden relationships between individual customer purchasing decisions [45]. The applicability and strength of such rules is assessed through their support and confidence, which measure the overall fraction of entities to which a particular rule applies, and the percentage of cases in which the co-occurrence can be observed, respectively.

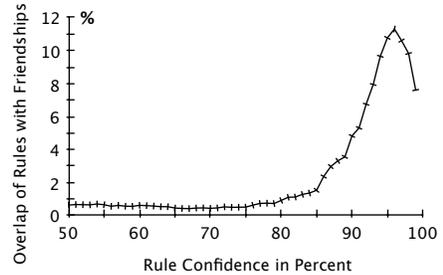

**Figure 13** Overlap of friendship topology with behavioral rules. When comparing the overlap in behavior between pairs of users (x-axis) and their likelihood to have a friendship relation (y-axis), it is found that the vast majority of persons who commonly click together on stories are not related at a topological level. The probability for two users A and B showing nearly identical behavioral patterns (95% of A's diggs are mirrored by B) to be friends is less than 12%.

We limited our search only to rules providing a minimum support of 0.01% and a minimum confidence of 50%, meaning that any user considered for a particular rule must have participated in at least 0.01% of all stories (thereby eliminating abandoned and very low-volume user accounts) and establishing only a relationship if users share at least half of their diggs together. For the entire corpus, nearly 1.2 million common activity patterns could be discovered, which were mapped against the topology of the actual friendship network.

Figure 13 shows the percentage of friendship links between user pairs that were found to exhibit high levels of co-participation on the same stories as a function of the rule confidence in percent. As can be seen from the figure, the vast majority of similarly behaving user pairs in the Digg network have not formed a friendship between them. For any confidence value between 50-80%, meaning that in 5-8 out of 10 cases a digg by user A on a particular story will result in a digg from user B, there is less than a 1% probability that user A and B are directly connected. Even for extremely high performing rules, when in 95 out of 100 cases two users behave in an identical manner, less than 12% of those user pairs are friends. We can therefore conclude that although there exist some patterns in the common behavior of users,

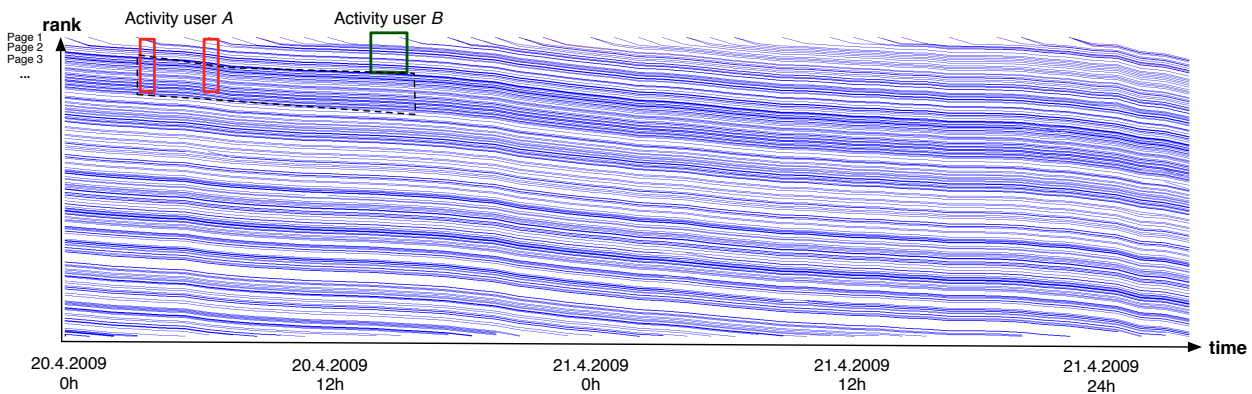

**Figure 12** Story placement on front pages over time. The figure shows the development of the absolute position of stories on the front pages (y-axis) as stories age and are displaced by newly promoted material over time, based on a 48-hour snapshot in April 2009 (x-axis). The color intensity of the line indicates the amount of diggs a story has currently accumulated. The high turnover rate of even the popular stories and the limited attention span and activity periods of users can offer an explanation of the low importance of the bare topology of the friendship graph is unsuited to fully capture it.

## 7.2 The "Digg Patriots": A Hidden Logical Layer

The existence of strong patterns and structure in the behavior of users on the website suggest that users may engage in community building, forming a "logical network topology" characterized by a specific semantics on top of the underlying social media platform. Concrete mechanisms to discover and identify the size and shape of these communities is still subject to research, as a proof-of-concept we will however make the case of the "Digg Patriots", an activist group of Digg users aiming to game the promotion algorithm through coordinated collective digging on stories, in some reported cases after payment by third parties [46].

When an email list archive of alleged members of the "Patriots" was exposed in 2010, we were able to link the email communications of 102 members to a particular Digg profile and cross-reference their identities against the discovered highly-aligned activity patterns of users. Nearly half of the exposed "Patriots" also appear in the body of discovered association rules, figure 14 shows the percentage of rules between 50-80% confidence from user interaction that were either linkable to either friendships or the "Digg Patriots". Remarkably, the collection of 102 coordinated "Patriots" known to us provides nearly a fifth of the discovered behavioral rules that can be extracted from the entire social network graph of 2 million users and 7.7 million friendship links between them.

It is therefore evident that effective social network analysis needs to go far beyond the analysis of the bare friendship topology and actually classify the semantics and characteristics of visible friendship and invisible other logical ties between social network users. How dramatic the logical ties between these particular users, undiscoverable from a graph theoretical perspective, might have been to the Digg social network can be exemplarily seen in figure 15 which shows the diggs made over time on three promoted stories: Coordinating and orchestrating diggs in the early life time of a story, the "Patriots" (indicated in red) might have been the driver influencing the trajectory enough to push them over the promotion threshold, thereby leveraging the mass attention resulting from a front page through very little effort, yet invisible topologically.

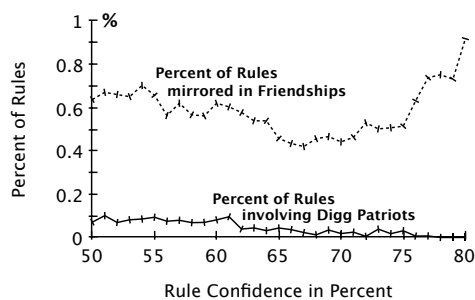

**Figure 14** Explanatory power of the "Digg Patriots" layer. Even though the 102 analyzed members of the "Digg Patriots" were only about a 20000[th] of the entire user population, this small select group – semantically and topologically invisibly linked by their group membership – alone generated a fifth of all common globally-recognizable activity patterns that could be found between the remaining 2 million users and their 7.7 million friendship links.

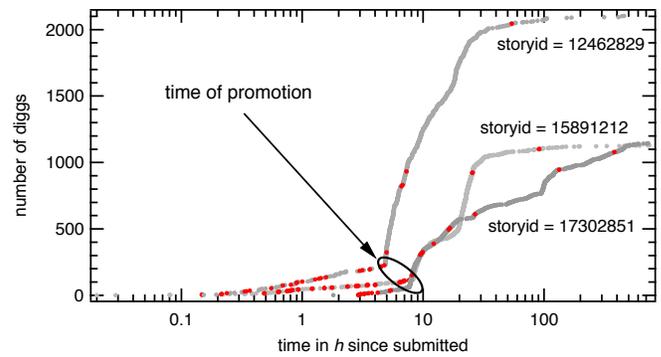

**Figure 15** Concentrated activities of "Digg Patriots". The figure shows the number of total diggs (y-axis) over time since the original submission of the story (x-axis), for the case of three example stories which became popular after unusually high digging activity from "Digg Patriots". The casting of an up-vote by a "Digg Patriot" is indicated in red; most of their activity is concentrated right after submission until the story reaches the promotion threshold and enters the front page.

# 8 Conclusion

In this paper, we have evaluated the common assumption made in social network analysis that the deciding factor determining whether some information goes viral or not are the individual friendship relations among users. While we find evidence of some structure in how these friendship relations are formed (there is a high overlap of interests), the actual effectiveness and common clicking rate of friendship links is surprisingly low and does not confirm the high importance that is attributed to these social ties. As the wider *network* of friends stretching over multiple hops (friends tell their friends who tell their friends) provides a much smaller contribution in practice than it could in theory (it could reach an exponentially increasing number of entities), the impact and the propagation along friendships and the network of friends is in most cases not enough to reach critical mass. We furthermore notice that although there exists a significant skew in the characteristics of network nodes from a topological perspective, we do not find any evidence that these network nodes are indeed behaving differently and more effectively in terms of spreading information. They have no better access to information, are not more efficient triggering their friends or do not predict trends better. The fact that there exists no group that can consistently and at a high level generate "hits" and individuals' success ratios fluctuate largely across observation periods leads to the conclusion that even successful members do not actually seem to have the recipe for success.

Various outcomes of our analysis however point to two factors that in the past have not received sufficient attention: time alignment and existence of non-topological relationships between users. We find that when incorporating these factors, the conductivity of information propagation and our ability to explain it in retrospect improves manyfold. Accurately predicting when users will be active and developing methods to detect and characterize these logical links between users will be the focus of future research.